\documentclass[aps,prd,preprint,showpacs,nofootinbib]{revtex4}
\usepackage{epsfig}
\usepackage{graphicx}
\usepackage{latexsym}
\usepackage{amssymb}
\usepackage{amsmath}

\newcommand{\msbis}{m^2_{\tilde{b}_i}}

\newcommand{\msbjs}{m^2_{\tilde{b}_j}}

\newcommand{\mstis}{m^2_{\tilde{t}_i}}

\newcommand{\mstjs}{m^2_{\tilde{t}_j}}

\newcommand{\mstks}{m^2_{\tilde{t}_k}}
\newcommand{\mgl}{m_{\tilde{g}}}
\newcommand{\mgls}{m^2_{\tilde{g}}}
\newcommand{\msqa}{m^2_{\tilde{q}_1}}
\newcommand{\msqb}{m^2_{\tilde{q}_2}}
\newcommand{\rqi}{\mbox{R}^{\tilde{q}}_{i1}}
\newcommand{\Rqi}{\mbox{R}^{\tilde{q}}_{i2}}
\newcommand{\rbi}{\mbox{R}^{\tilde{b}}_{i1}}
\newcommand{\Rbi}{\mbox{R}^{\tilde{b}}_{i2}}
\newcommand{\rbj}{\mbox{R}^{\tilde{b}}_{j1}}
\newcommand{\Rbj}{\mbox{R}^{\tilde{b}}_{j2}}

\newcommand{\rti}{\mbox{R}^{\tilde{t}}_{i1}}
\newcommand{\Rti}{\mbox{R}^{\tilde{t}}_{i2}}
\newcommand{\rtj}{\mbox{R}^{\tilde{t}}_{j1}}
\newcommand{\Rtj}{\mbox{R}^{\tilde{t}}_{j2}}
\newcommand{\rtk}{\mbox{R}^{\tilde{t}}_{k1}}

\newcommand{\sws}{\sin^2\theta_w}

\begin{document}

\title{ SUSY-QCD Corrections to $W^{\pm}H^{\mp}$ Associated Production at
the CERN Large Hadron Collider}
\author{\center{Jun Zhao, Chong Sheng Li\footnote{\hspace{-0.1cm}
Electronics address: csli@pku.edu.cn}, and Qiang Li }}
\affiliation{\small Department of Physics, Peking University,
Beijing 100871, China}

\begin{abstract}
We calculate the SUSY-QCD corrections to the inclusive total cross
sections of the associated production processes $pp\rightarrow
W^{\pm}H^{\mp}+X$ in the Minimal Supersymmetric Standard
Model(MSSM) at the CERN Large Hadron Collider(LHC). The SUSY-QCD
corrections can increase and decrease the total cross sections
depending on the choice of the SUSY parameters. When $\mu<0$ the
SUSY-QCD corrections increase the leading-order (LO) total cross
sections significantly for large tan$\beta$ ($\sim 40$), which can
exceed $10\%$ and have the opposite sign with respect to the QCD
and the SUSY-EW corrections, and thus cancel with them to some
extent. Moreover, we also investigate the effects of the SUSY-QCD
on the differential distribution of cross sections in transverse
momentum $p_T$ and rapidity Y of W-boson, and the invariant mass
$M_{W^+H^-}$.
\end{abstract}

\maketitle
\subsection{Introduction}
The Higgs mechanism\cite{higgs} plays a key role for spontaneous
breaking of the electroweak symmetry both in the standard
model(SM)and in the minimal supersymmetric extension of the
SM(MSSM)\cite{mssm}. The SM contains one neutral CP-even Higgs
boson, while the MSSM accommodates five physical Higgs bosons: the
neutral CP-even $h^0$ and $H^0$ bosons, the neutral CP-odd $A^0$
boson, and the charged $\mbox{H}^{\pm}$ boson pair. The charged
Higgs bosons do not belong to the spectrum of the SM, so the
discovery of them would be an unambiguous sign of new physics
beyond the SM. Therefore, the search of charged Higgs bosons
become one of the prime tasks in future high-energy experiments,
especially at the LHC\cite{higgssearchlhc}. At hadron colliders,
the charged Higgs bosons $\mbox{H}^{\pm}$ could appear as the
decay product of primarily produced top quarks if the mass of
$\mbox{H}^{\pm}$  is smaller than the $m_t-m_b$. For heavier
$\mbox{H}^{\pm}$, the main channels for single charged Higgs
production may be those associated with heavy quark, such as
$gb\rightarrow H^-t$\cite{gbthqcd} and $qb\rightarrow
q^{\prime}bH^-$\cite{qbqcd}. Although these processes give rather
large production rates, they suffer from also large QCD
backgrounds, especially when the $\mbox{H}^{\pm}$ mass is above
the threshold of $t\bar{b}$. And the channels for pair production
are $q\bar{q}$ annihilation and the loop-induced $gg$ fusion
processes \cite{higgspair}, which is also severely plagued by the
QCD backgrounds.

Another attractive mechanism of $\mbox{H}^{\pm}$ production in
association with $\mbox{W}^{\mp}$ bosons at hadron colliders has
been proposed and analyzed in the Ref.~\cite{whadvise}, which found
that the $W^{\mp}H^{\pm}$ production would have a sizeable cross
section and its signal should have a significant rate at the LHC
unless $m_{H^\pm}$ is very large. The dominant partonic subprocesses
of $W^{\mp}H^{\pm}$ associated production are $b\bar{b}\rightarrow
W^{\mp}H^{\pm}$ at the tree-level and $gg\rightarrow W^{\mp}H^{\pm}$
at the one-loop level. In these processes, the leptonic decays of
W-boson would serve as a spectacular trigger for the
$\mbox{H}^{\pm}$ search. A careful signal-versus-background analysis
has been done in Refs.\cite{whsignalvsbackground}. The detailed
computation of the cross section of the $gg$ fusion process can be
found in the Ref.~\cite{ggwh}. For the $b\bar{b}$ annihilation
process, the ${\cal O}(\alpha_{ew}m^2_{t(b)}/m^2_{W})$ and ${\cal
O}(\alpha_{ew}m^4_{t(b)}/m^4_{W})$ SUSY-EW and the ${\cal
O}(\alpha_s)$ QCD corrections also have been calculated in
Refs.~\cite{bbsusyew} and \cite{bbqcd}, respectively. However, the
one-loop SUSY-QCD corrections have not been reported in literatures
so far. So in this paper, we present the calculations of the
one-loop SUSY-QCD corrections to the process.

This paper is organized as follows. In Section B, we present some
analytic results for the cross sections of $b\bar{b}\rightarrow
W^+H^-$. In Section C, we give the numerical predictions for
inclusive and differential cross sections at the LHC. The relevant
SUSY Lagrangians and the lengthy analytic expressions are
summarized in the Appendices.

\subsection{Analytical Results}
We consider the associated production of $W^+H^-$ from the
collision of the two protons with momentum $P_1$ and $P_2$ at the
LHC. First, we define the Mandelstam variables of the subprocess
$b(p_1)\bar{b}(p_2)\rightarrow W^+(p_4)H^-(p_3)$ as
\begin{eqnarray}
&s=(p_1+p_2)^2=(p_3+p_4)^2, \nonumber\\
&t=(p_1-p_3)^2=(p_2-p_4)^2, \\
&u=(p_1-p_4)^2=(p_2-p_3)^2. \nonumber
\end{eqnarray}
Here the Mandelstam invariants are related by $s+t+u=m_W^2+m_H^2$.

We carry out the calculations in the t'Hooft-Feynman gauge and use
dimensional reduction, which preserves supersymmetry, to regularize
the UV divergences in the virtual loop corrections. In order to
remove the UV divergences, we renormalize the quark masses in the
Yukawa couplings and their wave functions by using the on-mass-shell
scheme\cite{onmassshell}. Denoting $m_{q0}$ and $\psi_{q0}$ as the
bare quark mass and the bare wave function, respectively, the
relevant renormalization constants $\delta m_q, \delta Z^q_{L}$ and
$\delta Z^q_{R}$ are then defined as
\begin{eqnarray}
&& m_{q0}=m_q +\delta m_q,\\
 && \psi_{q0}=(1+\delta Z^q_{L})^{\frac{1}{2}}\psi_{qL}+(1+\delta
 Z^q_{R})^{\frac{1}{2}}\psi_{qR}.
\end{eqnarray}
After calculating the self-energy diagram in Fig.~1, we obtain the
explicit expressions of all the renormalization constants as
follows:
\begin{eqnarray}
\delta
m_q=&&-\frac{\alpha_s}{3\pi}\big[m_q[B_1(m^2_q,\mgls,\msqa)+B_1(m^2_q,\mgls,\msqb)]+
            \mgl\sin2\theta_{\tilde{q}}[B_0(m^2_q,\mgls,\msqa) \\
            &&-B_0(m^2_q,\mgls,\msqb)]\big],  \nonumber  \\
\delta
Z^q_L=&&\frac{2\alpha_s}{3\pi}\big[\cos^2\theta_{\tilde{q}}B^1_1+\sin^2\theta_{\tilde{q}}B^2_1
             +m^2_q[\dot B^1_1+\dot B^2_1+\frac{\mgl}{m_q}\sin2\theta_{\tilde{q}}(\dot B^1_0-\dot B^2_0)]\big], \\
\delta
Z^q_R=&&\frac{2\alpha_s}{3\pi}\big[\sin^2\theta_{\tilde{q}}B^1_1+\cos^2\theta_{\tilde{q}}B^2_1
             +m^2_q[\dot B^1_1+\dot B^2_1+\frac{\mgl}{m_q}\sin2\theta_{\tilde{q}}(\dot B^1_0-\dot
             B^2_0)]\big],
\end{eqnarray}
where $B^i_m=B_m(m^2_q,\mgls,m^2_{\tilde{q}_i}), \dot B^i_m=\dot
B_m(m^2_q,\mgls,m^2_{\tilde{q}_i})$ with $q=b,\, t$ are the
two-point integrals \cite{denner}, $m_{\tilde{q}_{i}}$ are the
squark masses, $m_{\tilde{g}}$ is the gluino mass, and
$\theta_{\tilde{q}}$ is the mixing angle of the squarks.

The Feynman diagrams for the subprocess $b\bar{b}\rightarrow
H^{-}W^{+}$, which include the SUSY-QCD corrections to the
process, are shown in Fig.1 and its renomalized amplitude can be
written as
\begin{eqnarray}
M^{ren.}=M^0+M^{vir.}+\delta M^{count.}.
\end{eqnarray}
Here $M^0$ is the tree-level amplitude, which is given by summing
over the s- and t-channel amplitudes:
\begin{eqnarray}
 M^0= M^{s0}+ M^{t0},
\end{eqnarray}
with
\begin{eqnarray}
&& M^{s0}=\frac{2\pi\alpha
m_b}{m_w\sin^2\theta_w}\bigg[S_b(A_4+A_3)+P_b(A_4-A_3)\bigg],\\
&& M^{t0}=\frac{2\pi\alpha
}{m_w\sin^2\theta_w}\frac{1}{t-m_t^2}\big[m^2_t\cot\beta
A_{10}+m_b\tan\beta(A_{11}-2A_3)\big],
\end{eqnarray}
where $S_b$ and $P_b$ are defined by
\begin{eqnarray}
&&S_b=\frac{1}{\cos\beta}\left[-\frac{\sin\alpha\cos(\alpha-\beta)}{s-m^2_{h^0}}+\frac{\cos\alpha\sin(\alpha-\beta)}{s-m^2_{H^0}}\right],\\
&&P_b=\frac{1}{\cos\beta}\frac{\sin\beta}{s-m^2_{A^0}},
\end{eqnarray}
$A_i$ are reduced standard matrix elements, which are given by
\begin{eqnarray}
\label{me}
&&A_1= \bar v(p_2) P_R u(p_1)p_1\cdot\varepsilon(p_4),\nonumber\\
&&A_2= \bar v(p_2) P_L u(p_1)p_1\cdot\varepsilon(p_4),\nonumber\\
&&A_3= \bar v(p_2) P_R u(p_1)p_3\cdot\varepsilon(p_4),\nonumber\\
&&A_4= \bar v(p_2) P_L u(p_1)p_3\cdot\varepsilon(p_4),\\
&&A_5= \bar v(p_2)\not p_3P_R u(p_1)p_1\cdot\varepsilon(p_4),\nonumber\\
&&A_6= \bar v(p_2)\not p_3P_L u(p_1)p_1\cdot\varepsilon(p_4),\nonumber\\
&&A_7= \bar v(p_2)\not p_3P_R
u(p_1)p_3\cdot\varepsilon(p_4),\nonumber
\end{eqnarray}
\begin{eqnarray}
&&A_8= \bar v(p_2)\not p_3P_L u(p_1)p_3\cdot\varepsilon(p_4),\nonumber\\
&&A_9= \bar v(p_2)\not\varepsilon(p_4)P_Ru(p_1),\nonumber\\
&&A_{10}= \bar v(p_2)\not\varepsilon(p_4)P_L u(p_1),\nonumber\\
&&A_{11}= \bar v(p_2)\not p_3\not\varepsilon(p_4)P_Ru(p_1),\nonumber\\
&&A_{12}= \bar v(p_2)\not
p_3\not\varepsilon(p_4)P_Lu(p_1),\nonumber
\end{eqnarray}
and $M^{vir.}$ contains the radiative corrections from the one-loop
self-energy, vertex, and box diagrams, of which corresponding
amplitudes are shown in Appendix B. The counter-term $\delta
M^{count.}$ contains the contributions from the corresponding vertex
and self-energy counterterms, which are given by
\begin{eqnarray}
\delta M^{vs}=&&M^{s0}(\frac{\delta m_b}{m_b}+\frac{1}{2}\delta Z^b_R+\frac{1}{2}\delta Z^b_L),\\
\delta M^{vt1}=&&\frac{2\pi\alpha
}{m_w\sin^2\theta_w}\frac{1}{t-m_t^2}\big[m^2_t\cot\beta(\frac{\delta
m_t}{m_t}+\frac{1}{2}\delta Z^t_R+\frac{1}{2}\delta Z^b_L) A_{10} \nonumber\\
           &&+m_b\tan\beta(\frac{\delta m_b}{m_b}+\frac{1}{2}\delta Z^t_L+\frac{1}{2}\delta Z^b_R)(A_{11}-2A_3)\big],\\
\delta M^{vt2}=&&M^{t0}(\frac{1}{2}\delta Z^b_L+\frac{1}{2}\delta Z^t_L),\\
\delta
M^{self}=&&\frac{2\pi\alpha}{m_w\sin^2\theta_w}\frac{1}{t-m^2_t}\bigg[m^2_t\cot\beta(\frac{\delta
m_t}{m_t}
                +\frac{1}{2}\delta Z^t_L+\frac{1}{2}\delta
                Z^t_R)A_{10} \nonumber \\&&
                +m_b\tan\beta(\delta Z^t_L+\frac{2m^2_t}{t-m^2_t}\frac{\delta m_t}{m_t})(A_{11}-2A_3)\bigg].
\end{eqnarray}
The partonic cross section can be written as following:
\begin{eqnarray}
\hat{\sigma}=\int^1_{-1}\mbox{d}z\frac{1}{32\pi
s^2}\lambda^{1/2}\overline{\left|M^{ren.}\right|^2}=\int^{t_+}_{t_-}\mbox{d}t\frac{1}{16\pi
                     s^2}\overline{\left|M^{ren.}\right|^2},
\end{eqnarray}
where $\lambda\equiv(s+m^2_W-m^2_{H^-})^2-4sm^2_W$,
$t_{\pm}=\frac{1}{2}[m^2_{H^-}+m_W^2-s\pm\lambda^{1/2}],$ and
$\overline{\left|M^{ren.}\right|^2}$ is the renormalized amplitude
squared, which is given by
\begin{eqnarray}
\overline{\left|M^{ren.}\right|^2}=\overline\sum\left|M^0\right|^2+2\mbox{Re}\overline\sum
M^0[M^{vir.}+\delta M^{count.}]^{\dagger},
\end{eqnarray}
where the colors and spins of the outgoing particles have been
summed over, and the colors and spins of incoming ones have been
averaged over. We notice that the color average factors of LO and
NLO amplitudes squared are different:
$\frac{1}{3}\times\frac{1}{3}\mbox{Tr}{\openone }=\frac{1}{3}$ for
LO amplitudes, and
$\frac{1}{3}\times\frac{1}{3}\mbox{Tr}(\mbox{T}^a\mbox{T}^a)=\frac{4}{9}$
for NLO ones. The both spin average factors are the same
$\frac{1}{2}\times\frac{1}{2}=\frac{1}{4}$.

The total cross section at the LHC is obtained by convoluting the
partonic cross section with the parton distribution functions
(PDFs) $G_{b,\bar{b}/p}$ in the proton:
\begin{eqnarray}
\sigma=\int^1_{\tau_0} dx_1\int^1_{\tau_0/x_1} dx_2
[G_{b/p}(x_1,\mu_f)G_{\bar{b}/p}(x_2,\mu_f)+ (x_1\leftrightarrow
x_2)]\hat{\sigma}(\tau S),
\end{eqnarray}
where $\mu_f$ is the factorization scale and $S=(P_1+P_2)^2$,
$P_1$ and $P_2$ are the four-momentum of the incident hadrons,
$\tau_0=\frac{(m_W+m_{H^-})^2}{S}$, $\tau =x_1x_2$, and $x_1$ and
$x_2$ are the longitudinal momentum fractions of initial partons
in the hadrons.

In the following, we present the differential cross sections in
the transverse momentum $p_T$ and rapidity Y of the W-boson, and
the invariant mass $M_{W^+H^-}$, respectively. In the
center-of-mass frame of initial hadrons, $P_1=\sqrt{S}/2(1,0,0,1)$
and $P_2=\sqrt{S}/2(1,0,0,-1)$, and the four-momentum of W-boson
is defined by $p_4=(E,\mathbf{p_T},p_L)$. The transverse momentum
$p_T$ and the rapidity Y of W-boson, and the invariant mass
$M_{W^+H^-}$ are defined by
\begin{eqnarray}
p_T^2&=&(E-p_L)(E+p_L)-m_W^2=\frac{(m_W^2-t)(m^2_{W}-u)}{s}-m_W^2,\\
Y&=&\frac{1}{2}\log\left(\frac{E+p_L}{E-p_L}\right),
\end{eqnarray}
and
\begin{eqnarray}
\hspace{-2cm}M^2_{W^+H^-}=(p_3+p_4)^2=(p_1+p_2)^2=s=x_1x_2S,\end{eqnarray}
respectively.

The three differential cross sections are thus given by
\begin{eqnarray}
\frac{\mbox{d}{\sigma}}{\mbox{d}p_T}=\int^1_{\tau_0}
dx_1\int^1_{\tau_0/x_1} dx_2
[G_{b/p}(x_1,\mu_f)G_{\bar{b}/p}(x_2,\mu_f)+ (x_1\leftrightarrow
x_2)]\frac{\mbox{d}\hat{\sigma}}{\mbox{d}p_T}\label{pt},
\end{eqnarray}
\begin{eqnarray}
\frac{\mbox{d}{\sigma}}{\mbox{d}Y}=\int^1_{\tau_0}
dx_1\int^1_{\tau_0/x_1} dx_2
[G_{b/p}(x_1,\mu_f)G_{\bar{b}/p}(x_2,\mu_f)+ (x_1\leftrightarrow
x_2)]\frac{\mbox{d}\hat{\sigma}}{\mbox{d}Y} \label{rapi},
\end{eqnarray}
and
\begin{eqnarray}
\frac{\mbox{d}{\sigma}}{\mbox{d}M_{W^+H^-}}=\int^1_{\tau_0}\frac{\mbox{d}x_1}{x_1}\frac{2M_{W^+H^-}}{\tau
S}[G_{b/p}(x_1,\mu_f)G_{\bar{b}/p}(x_2,\mu_f)+ (x_1\leftrightarrow
x_2)]\mbox{d}\hat{\sigma},\label{inm}
\end{eqnarray}
respectively,\\
with
\begin{eqnarray}
\frac{\mbox{d}\hat{\sigma}}{\mbox{d}p_T}=\frac{1}{32\pi
s^2}\lambda^{1/2}\lvert\frac{4sp_T}{z\lambda}\rvert\overline{\left|M^{ren.}\right|^2},
\end{eqnarray}
and \begin{eqnarray}
\frac{\mbox{d}\hat{\sigma}}{\mbox{d}Y}=\frac{1}{16\pi
s^2}\lvert(s+m_W^2-m^2_{H^-})\frac{2u}{(u+1)^2}\rvert\overline{\left|M^{ren.}\right|^2},
\end{eqnarray}
where $z=\pm \sqrt{1-\frac{4sp^2_T}{\lambda}}$ and
$u=\frac{x_1}{x_2}e^{-2Y}$.

\subsection{Numerical Results and Conclusions}
In this section, we present the numerical results for the SUSY-QCD
corrections to $W^+H^-$ associated production at the LHC. In our
numerical calculations, we used the following set of the SM
parameters\cite{smparameter}:
\begin{eqnarray}
&& \alpha_{ew}(m_W)=1/128, \ m_W=80.419 \, {\rm \,GeV},
\ m_Z=91.1882 \,{\rm \,GeV} \, ,\nonumber\\
&& m_b=4.25\,{\rm \,GeV},m_t=178 \, {\rm \,GeV}, \
\alpha_s(M_Z)=0.118.
\end{eqnarray}

The running QCD coupling $\alpha_s(Q)$ is evaluated at the
two-loop order~\cite{2loopalfs}, and the CTEQ6M PDFs \cite{pdf}
are used throughout this paper either at the LO or NLO. For
simplicity, we neglect the b-quark mass but keep it in the
couplings. Moreover, in order to improve the perturbative
calculations, we took the running mass $m_b(Q)$ and $m_t(Q)$
evaluated the NLO formula\cite{mrunning}:
\begin{eqnarray}
&&m_b(Q)=U_6(Q,m_t)U_5(m_t,m_b)m_b(m_b) \, ,\\
&&m_t(Q)=U_6(Q,m_t)m_t(m_t) \, ,
\end{eqnarray}
where the evolution factor $U_f$ is
\begin{eqnarray}
U_f(Q_2,Q_1)=\bigg(\frac{\alpha_s(Q_2)}{\alpha_s(Q_1)}\bigg)^{d^{(f)}}
\bigg[1+\frac{\alpha_s(Q_1)-\alpha_s(Q_2)}{4\pi}J^{(f)}\bigg], \nonumber \\
d^{(f)}=\frac{12}{33-2f}, \hspace{1.0cm}
J^{(f)}=-\frac{8982-504f+40f^2}{3(33-2f)^2} \, ,
\end{eqnarray}
and $f$ is the number of the active light quarks.

In addition, to also improve the perturbation calculations, we
made the following SUSY replacements in the tree-level
couplings\cite{mrunning,mtrunning}
\begin{eqnarray}
&& m_q(Q) \ \ \rightarrow \ \ \frac{m_q(Q)}{1+\Delta m_q}
\label{deltamb}\,\,\,\,(q=b,t),
\end{eqnarray}
\begin{eqnarray}
\Delta m_b=&&\frac{2\alpha_s}{3\pi}M_{\tilde{g}}\mu\tan\beta I(m_{\tilde{b}_1},m_{\tilde{b}_2},M_{\tilde{g}})\label{deltamb1}\, ,\\
\Delta m_t=&&-\frac{\alpha_s}{3\pi}\{\bar{B}_1(0,m^2_{\tilde{g}},m^2_{\tilde{t}_1})+\bar{B}_1(0,m^2_{\tilde{g}},m^2_{\tilde{t}_2})\nonumber\\
&&-\sin2\theta_t(\frac{m_{\tilde{g}}}{m_t})[\bar{B}_0(0,m^2_{\tilde{g}},m^2_{\tilde{t}_1})-\bar{B}_0(0,m^2_{\tilde{g}},m^2_{\tilde{t}_1})]\},
\end{eqnarray}
where
\begin{eqnarray}
I(a,b,c)=\frac{1}{(a^2-b^2)(b^2-c^2)(a^2-c^2)}
(a^2b^2\log\frac{a^2}{b^2} +b^2c^2\log\frac{b^2}{c^2}
+c^2a^2\log\frac{c^2}{a^2}) \, ,
\end{eqnarray}
$$\bar{B}_1=B_1+\Delta/2,\bar{B}_0=B_0-\Delta
\,\,\,(\Delta=1/\epsilon-\gamma+\ln4\pi).$$
It is necessary to
avoid double counting by subtracting these SUSY-QCD corrections
from the renormalization constant. As for the renormalization and
factorization scales, we always chose
$\mu_r=\mu_f=(m_W+m_{H^-})/2$.

The values of the MSSM parameters taken in our numerical
calculations were constrained within the minimal supergravity
scenario (mSUGRA)\cite{msugra}, in which there are only five free
input parameters at the grand unification where $m_{1/2},\, m_0,\,
A_0, \mbox{tan}\beta$ and the sign of $\mu$, where $m_{1/2},
\,m_0, \,A_0$ are, respectively, the universal gaugino mass,
scalar mass, and the trilinear soft breaking parameter in the
superpotential. Given these parameters, all the MSSM parameters at
the weak scale are determined in the mSUGRA scenario by using the
the program package SUSPECT 2.3\cite{suspect}.

Figs.~2 and ~3 show the total cross sections and the relative
corrections as functions of $m_{1/2}$ (or $m_{\tilde{g}}$) for
$\mu < 0$ and tan$\beta$=4,\, 10,\, and 40, respectively. The
total cross sections decrease with the increasing of $m_{1/2}$ as
expected. For small tan$\beta$, the relative SUSY-QCD corrections
are small and can be negligible. For large tan$\beta (\sim 40)$,
the relative SUSY-QCD corrections become large and increase when
$m_{\tilde{g}}$ increases. Indeed, in our numerical calculations,
the dependence of the total cross sections on $m_{\tilde{g}}$ is
got through varying $m_{1/2}$. And when $m_{1/2}$ increases, both
$m_{\tilde{g}}$ and $m_{H^{\pm}}$ increase. The increase of
$m_{H^{\pm}}$ decreases the phase spaces of the LO total cross
sections, and the magnitudes of the SUSY-QCD corrections also
become smaller with the increasing of both $m_{\tilde{g}}$ and
$m_{H^{\pm}}$. But the decrease rate of the LO total cross
sections is larger than the one of the SUSY-QCD corrections, so
the relative corrections increase with the increasing of $m_{1/2}$
as shown in Fig.~3.

In Figs.~4 and ~5 we present the LO and the SUSY-QCD corrected
total cross sections, and the SUSY-QCD corrections as functions of
$m_{H^-}$ for $\mu<0$ and $\tan\beta=4, 10, 40$, respectively.
Both the LO and the SUSY-QCD corrected cross sections decrease
when $m_{H^-}$ increases, and for large $\tan\beta$, the
corrections in general enhance the total cross sections for
$\mu<0$. For large $\mbox{tan}\beta (\sim 40)$ the total cross
sections become significant and can reach several tens fb, and
even one hundred fb when $m_{H^-}<200\mbox{\,GeV}$, while for
small $\mbox{tan}\beta$ the total cross sections are about serval
fb and can be neglected. For $\mbox{tan}\beta=40$, in general, the
corrections can exceed $4\%$, and even they can reach $10\%$ when
$m_{H^-}\sim 150\mbox{\,GeV}$. For $\tan\beta=4, 10$, the
magnitude of the corrections are always smaller than $2 \%$. Note
that for $\mu <0$ and $\tan\beta=4, 10$, the mass of the charged
Higgs can not be smaller than about $250$ GeV, just as shown by
the curves in these figures.

Figs.~6 and ~7 show the cross sections and the relative
corrections as functions of tan$\beta$, assuming $m_0=160,\,400$
GeV, and the sign of $\mu=\pm 1$, respectively. From these
figures, we find that the total cross sections for the case of
$\mu > 0$ are obviously smaller than those for $\mu <0$, and that
when tan$\beta$ becomes larger the SUSY-QCD corrections increase
the total cross sections for $\mu <0$, while decrease for $\mu
>0$. For small $\tan\beta$, the variations of these curves are not
monotonic because the contributions from the Yukawa coupling
$\mbox{H}^-t\bar{b}$ contain not only the terms proportional to
tan$\beta$ but also the ones to cot$\beta$.

Above results for representative values of $m_{H^-}$ and
tan$\beta$ can be summarized in table~\ref{tanbeta}.
\begin{table}[t]
\begin{tabular}{|c|c|c|c|}
  \hline
   $m_{H^+}$ GeV & \ \ tan$\beta$=4 \ \ \ & \ \ tan$\beta$=10 \ \ \ & tan$\beta$=40 \\
  \hline
   150 \ \  & \ \ $-$ \ \ & \ \ $-$ \ \  & \ \ $\sim$ 10 $\%$ \ \ \\
  \hline
  300 \ \  & \ \ $\sim 0.2 \%$ \ \ & \ \ $\sim - 0.2 \%$ \ \  & \ \ $\sim 5 \%$ \ \ \\
  \hline
  500 \ \  & \ \ $\sim 0.8 \%$ \ \ & \ \ $\sim - 0.7 \%$ \ \  & \ \ $\sim 4 \%$ \ \ \\
  \hline
 \end{tabular}\caption{The SUSY-QCD corrections for tan$\beta$=4,\,
 10,\,40 and $m_{H^-}=150,\,300,\,500$ GeV,
respectively, assuming $A_0=250$ GeV, $m_{1/2}=180$ GeV, $\mu <
0$.}\label{tanbeta}
\end{table}

In Figs.8-10, we display the differential cross sections as
functions of the transverse momentum $p_T$, the rapidity Y of the
W-boson, and the invariant mass $M_{W^+H^-}$, which are given by
Eqs.\ref{pt}, \ref{rapi}, and \ref{inm}, respectively. We find that
the SUSY-QCD corrections increase the LO differential cross sections
for $\mu<0$, and decrease ones for $\mu>0$. The differential cross
sections can reach the maximum value at $p_T=55$\,GeV and 70\, GeV,
$\mbox{M}_{W^+H^-}=310$ and 370 \,GeV for $\mu <0 $ and $\mu >0$,
respectively. The differential cross sections in the rapidity of
W-boson is symmetric about the axis of Y=0 as expected.

Finally, we compare the SUSY-QCD corrections with the ${\cal
O}(\alpha_{ew}m^2_{t(b)}/m^2_{W})$  and ${\cal
O}(\alpha_{ew}m^4_{t(b)}/m^4_{W})$ SUSY-EW and the ${\cal
O}(\alpha_s)$ QCD corrections. We notice that the QCD and the
SUSY-EW corrections are large and dominate over the SUSY-QCD ones
for small $\tan\beta$. When $\tan\beta$ becomes large the SUSY-QCD
corrections increase. Although the magnitudes of the SUSY-QCD
corrections are still smaller than the ones of the ${\cal
O}(\alpha_s)$ QCD corrections, they can exceed $10\%$, which are
larger than those of the SUSY-EW corrections. Especially, for
$\mu<0$, the sign of SUSY-QCD corrections is opposite to the ones
of the other two corrections, thus the SUSY-QCD corrections can
cancel with the SUSY-EW and QCD corrections to some extent. In
order to compare these corrections clearly, we show the three
corrections to the process $b\bar{b}\rightarrow W^{-}H^{+}$ in
some typical parameter space in the table~\ref{compare}.

In conclusion, we have calculated the SUSY-QCD corrections to the
total cross sections for the $W^+H^-$ associated production in the
MSSM at the LHC. The SUSY-QCD corrections can increase and
decrease the total cross sections depending on the choice of the
SUSY parameters. For $\mu<0$, the SUSY-QCD corrections can
increase the total cross sections significantly, especially for
large tan$\beta$, which have the opposite sign with respect to the
QCD and the SUSY-EW corrections, and cancel with them to some
extent.
\begin{table}[t]
\begin{tabular}{|c|c|c|c|}
  \hline
  & \ \ ${\cal O}(\alpha_{ew}m^2_{t(b)}/m^2_{W})$ and ${\cal O}(\alpha_{ew}m^4_{t(b)}/m^4_{W})$  \ \ \ & \ \ ${\cal O}(\alpha_s)$ QCD  \ \ \ & SUSY-QCD \\
\raisebox{0.3cm}{tan$\beta$} \ \ & SUSY-EW corrections
\cite{bbsusyew}\ \
& corrections \cite{bbqcd}\ \ & corrections \ \ \\
  \hline
   4   & \ \ $\sim -14 \%$  \ \ & \ \ $\sim - 32\%$ \ \  & \ \ $\sim$ 0 \ \ \\
  \hline
  40   &  \ \  $\sim -1 \%$ \ \ & \ \ $\sim - 17 \%$ \ \  & \ \ $\sim 8 \%$ \ \ \\
  \hline
 \end{tabular}\caption{Comparison of the SUSY-QCD corrections with the QCD and SUSY-EW ones,
  for tan$\beta=4,\,40$, respectively, assuming $m_{H^-}=200$GeV and $\mu < 0$.}\label{compare}
  \end{table}

\section{Acknowledgements}
This work was supported in part by the National Natural Science
Foundation of China, under grant Nos.10421003 and 10575001, and
the Key Grant Project of Chinese Ministry of Education, under
grant NO.305001.
\section*{Appendix A}
In this Appendix, we will list the relevant pieces of SUSY
Lagranian. The Yukawa couplings of Higgs and quarks are given by
\begin{eqnarray*}
&&{\cal L}_{(h^0,H^0,A^0)\bar{b}b}=\frac{gm_b}{2m_w\cos\beta}\sin\alpha h^0\bar{b}b-\frac{gm_b}{2m_w\cos\beta}
                       \cos\alpha H^0\bar{b}b+i\frac{gm_b}{2m_w}\tan\beta A^0\bar{b}\gamma_5 b,\\
\vspace{0.5cm} &&{\cal
L}_{H^-\bar{t}b}=\frac{g}{\sqrt{2}m_w}H^+\bar{t}(m_t\cot\beta
P_{L}+m_b\tan\beta P_{R})b+h.c.,
\end{eqnarray*}
where $P_{L,R}=(1\mp \gamma_5)/2$ are the chiral projector
operators, tan$\beta=v_u/v_d$ is the ratio of vaccum expectation
values of the two Higgs doublets.

The trilinear couplings of Higgs bosons and W-boson are given by
\begin{eqnarray*}
&&{\cal
L}_{(h^0,H^0,A^0)H^-W^+}=i\frac{g}{2}\big[\cos(\alpha-\beta)h^0\overleftrightarrow{\partial_{\mu}}H^-
                              +\sin(\alpha-\beta)H^0\overleftrightarrow{\partial_{\mu}}H^-
                              +iA^0\overleftrightarrow{\partial_{\mu}}H^-\big]W^{+\mu}+h.c.
\end{eqnarray*}
The squarks couplings to gluino, W-boson and Higgs are given by
\begin{eqnarray*}
&&{\cal L}_{\tilde{g}q\tilde{q}}=-\sqrt{2}T^a[\bar{q}(\rqi
P_R-\Rqi P_L)\tilde{g}\tilde{q}_i
                               +\bar{\tilde{g}}(\rqi P_R-\Rqi
                               P_L)q\,\tilde{q}^{\ast}_i],\\
&&{\cal L}_{\tilde{t}^{\ast}\tilde{b}W}=i\frac{g}{\sqrt{2}}(\rbi\rtj W^{+\mu}\tilde{t}^{\ast}_j\overleftrightarrow{\partial}_{\mu}\tilde{b}_i
                              +\rti\rbj W^{-\mu}\tilde{b}^{\ast}_j \overleftrightarrow{\partial}_{\mu}\tilde{t}_i  ),\\
&&{\cal L}_{H^k\tilde{b}^{\ast}\tilde{b}}=gG^{k}_{ij}H^k\tilde{b}^{\ast}_{j}\tilde{b}_i \hspace{3cm} (H^k=h^0,H^0,A^0),\\
&&{\cal L}_{H^k\tilde{t}^{\ast}\tilde{b}}=\frac{g}{\sqrt{2}m_w}G^{k}_{ij}H^k\tilde{t}^{\ast}_{j}\tilde{b}_i  +h.c.\hspace{2cm}(H^k=H^-),\\
\end{eqnarray*}
where
$$\hspace{-2cm}\hspace{3cm}G^k_{ij}=[\mbox{R}^{\tilde{b}}\hat{G}^k\mbox{R}^{\tilde{b}T}]_{ij},$$
\begin{eqnarray*}
\hspace{-0.2cm}\hat{G}^{h^0}=\left(\begin{array}{cc}-\frac{m_z}{\cos\theta_w}(\frac{1}{2}-\frac{1}{3}\sin^2\theta_w)\sin(\alpha+\beta)
                                 +\frac{m^2_b}{2m_w\cos\beta}\sin\alpha  & \frac{m_b}{2m_w\cos\beta}(A_b\sin\alpha+\mu\cos\alpha)\\
                                  \frac{m_b}{2m_w\cos\beta}(A_b\sin\alpha+\mu\cos\alpha)&
                                  -\frac{1}{3}\frac{m_z}{\cos\theta_w}\sws\sin(\alpha+\beta)+\frac{m^2_b}{2m_w\cos\beta}\sin\alpha
\end{array}\right),
\end{eqnarray*}
\begin{eqnarray*}
\hspace{-0.2cm}\hat{G}^{H^0}=\left(\begin{array}{cc}\frac{m_z}{\cos\theta_w}(\frac{1}{2}-\frac{1}{3}\sws)\cos(\alpha+\beta)
                                 -\frac{m^2_b}{2m_w\cos\beta}\sin\alpha  & -\frac{m_b}{2m_w\cos\beta}(A_b\cos\alpha-\mu\sin\alpha)\\
                                  -\frac{m_b}{2m_w\cos\beta}(A_b\cos\alpha-\mu\sin\alpha)&
                                  \frac{1}{3}\frac{m_z}{\cos\theta_w}\sws\sin(\alpha+\beta)-\frac{m^2_b}{2m_w\cos\beta}\sin\alpha
\end{array}\right),
\end{eqnarray*}
\begin{eqnarray*}
\hspace{-7.5cm}\hat{G}^{A^0}=\frac{im_b}{2m_w}\left(\begin{array}{cc}0
&-(A_b\tan\beta+\mu)\\(A_b\tan\beta+\mu) & 0
\end{array}\right),
\end{eqnarray*}
\begin{eqnarray*}
\hspace{-5cm}\hat{G}^{H^-}=\left(\begin{array}{cc}m^2_b\tan\beta+m^2_t\cot\beta-m^2_w\sin2\beta
 &  m_b(A_b\tan\beta+\mu) \\
m_t(A_t\cot\beta+\mu) & \frac{2m_tm_b}{\sin2\beta}
\end{array}\right),
\end{eqnarray*}
where $R^{\tilde q}$ is a $2\times 2$ matrix shown as below, which
is defined to transform the squark current eigenstates to the mass
eigenstates \cite{squarkrotation}:
\begin{equation*}
\left(\begin{array}{c} \tilde{q}_1 \\ \tilde{q}_2 \end{array}
\right)= R^{\tilde{q}}\left(\begin{array}{c} \tilde{q}_L \\
\tilde{q}_R \end{array} \right), \ \ \ \ \
R^{\tilde{q}}=\left(\begin{array}{cc} \cos\theta_{\tilde{q}} &
\sin\theta_{\tilde{q}} \\ -\sin\theta_{\tilde{q}} &
\cos\theta_{\tilde{q}}
\end{array} \right),
\end{equation*}
with $0 \leq \theta_{\tilde{b}} < \pi$, by convention.
Correspondingly, the mass eigenvalues $m_{\tilde{q}_1}$ and
$m_{\tilde{q}_2}$ (with $m_{\tilde{q}_1}\leq m_{\tilde{q}_2}$) are
given by
\begin{eqnarray*}\label{Mq2}
\left(\begin{array}{cc} m_{\tilde{q}_1}^2 & 0 \\ 0 &
m_{\tilde{q}_2}^2 \end{array} \right)=R^{\tilde{q}}
M_{\tilde{q}}^2 (R^{\tilde{q}})^\dag, \ \ \ \ \
M_{\tilde{q}}^2=\left(\begin{array}{cc} m_{\tilde{q}_L}^2 & a_qm_q
\\ a_qm_q & m_{\tilde{q}_R}^2 \end{array} \right),
\end{eqnarray*}
with
\begin{eqnarray*}
m^2_{\tilde{q}_L} &=& M^2_{\tilde{Q}} +m_q^2
+m_Z^2\cos2\beta(I_{3L}^q -e_q\sin^2\theta_W), \\
m^2_{\tilde{q}_R} &=& M^2_{\tilde{D}} +m_q^2
+m_Z^2\cos2\beta e_q\sin^2\theta_W, \\
a_q &=& A_q -\mu\tan\beta.
\end{eqnarray*}
Here, $M_{\tilde{q}}^2$ is the squark mass matrix.
$M_{\tilde{Q},\tilde{D}}$ and $A_{q}$ are soft SUSY breaking
parameters and $\mu$ is the higgsino mass parameter. $I_{3L}^q$
and $e_q$ are the third component of the weak isospin and the
electric charge of the quark $q$, respectively.
\section*{Appendix B}
In this Appendix, we will list the explicit expressions of the
vertex, box and self-energy diagrams. For simplicity, we introduce
the following abbreviations for the Passarino-Veltman two-point
integrals $B_i$, tree-point integrals $C_{i(j)}$ and four-point
integrals $D_{ij}$, which are defined similarly to
Ref.~\cite{denner} except that we take internal masses squared as
arguments:
\begin{eqnarray*}
&&B_{i}=B_{i}(t,\mgls,\mstis),\\
&&C^a_{0}=C_{0}(0,s,0,\mgls,\msbis,\msbjs),\\
&&C^b_{i}=C_{i}(0,m^2_{H^-},t,\mgls,\msbis,\mstjs),\\
&&C^d_{i(j)}=C_{i(j)}(0,m^2_{w},t,\mgls,\msbis,\mstjs),\\
&&D_{ij}=D_{ij}(0,s,m^2_{w},t,0,m^2_{H^-},\mgls,\msbis,\msbjs,\mstks).
\end{eqnarray*}
The explicit expressions of the corresponding self-energy,vertex
and box diagrams are given by
\begin{eqnarray*}
&&{\cal
M}^{vs}=-\frac{2\alpha\alpha_s\mgl}{\sws}\bigg[\frac{G^{h^0}_{ij}\cos(\alpha-\beta)}{s-m^2_{h^0}}
                    +\frac{G^{H^0}_{ij}\sin(\alpha-\beta)}{s-m^2_{H^0}}+\frac{G^{A^0}_{ij}}{s-m^2_{A^0}}\bigg]
                    \big[\Rbi\rbj A_3+\rbi\Rbj A_4\big]C^a_0,\\
&&{\cal M}^{vt1}=\frac{\alpha\alpha_s}{\sws
m_w}\frac{G^{H^-}_{ij}}{t-m^2_t}\bigg[\big[\rbi\rtj t
C^b_2+\rbi\Rtj \mgl m_t C^b_0\big]A_{10}\\
     &&\hspace{1.5cm}+\big[\Rbi\rtj \mgl C^b_0+\Rbi\Rtj m_t C^b_2\big]\big[A_{11}-2
     A_3\big]\bigg],\\
&&{\cal M}^{vt2}=\frac{2\alpha\alpha_s}{\sws
m_w}\frac{\rbi\rtj}{t-m^2_t}\bigg[\rbi\rtj\big[m^2_t\cot\beta[C^d_{00}A_{10}+C^d_{12}(A_6-A_8)]
                 +m_b\tan\beta[C^d_{00}(A_{11}-2A_3)\\
&&\hspace{1.5cm} +t C^d_{12}(A_3-A_1)]\big]+\rbi\Rtj \mgl m_tC^d_1[\cot\beta(A_6-A_8)+m_b\tan\beta(A_3-A_1)]\\
&&\hspace{1.5cm}-\Rbi\rtj\mgl C^d_1\big[m^2_t\cot\beta(A_2-A_4)+m_b\tan\beta(A_7-A_5)\big]\\
&&\hspace{1.5cm}+\Rbi\Rtj
m_t\big[\cot\beta[C^d_{00}(A_{12}-2A_4)+tC^d_{12}(A_{4}-A_2)]\\
&&\hspace{1.5cm}+m_b\tan\beta[C^d_{12}(A_5-A_7)+C^d_{00}A_9]\big] \bigg],\\
&&{\cal M}^{box}=\frac{2\alpha\alpha_s}{\sws m_w}\rbj\rtk
                  G^{H^-}_{ki}\bigg[\rbi\rbj\big[A_{10}D_{00}-A_6(D_{13}+D_{23}+D_{33})
                 +A_8(D_{23}+D_{33}\big])\\
&&\hspace{1.5cm}+\Rbi\Rbj\big[A_9D_{00}-A_5(D_{13}+D_{23}+D_{33})+A_7(D_{23}+D_{33})\big]\bigg],\\
&&{\cal
       M}^{self}=\frac{\alpha\alpha_s}{\sws}\frac{1}{(t-m^2_t)^2}\bigg[m_t\cot\beta
\big[\mgl(m^2_{H^-}+m^2_t)\rti\Rti B_0
                 +m^2_{H^-}m_tB_1\big]A_{10}\\
&&\hspace{1.5cm} +m_b\tan\beta\big[m^2_{H^-}(\rti)^2+2\mgl
m_t\rti\Rti B_0+m^2_t(\Rti)^2B_1 \big][A_{11}-2A_3]\bigg],
\end{eqnarray*}
where we omit the common color factor $\mbox{T}^a\mbox{T}^a$.

\begin{figure}[h!]
\centerline{\epsfig{file=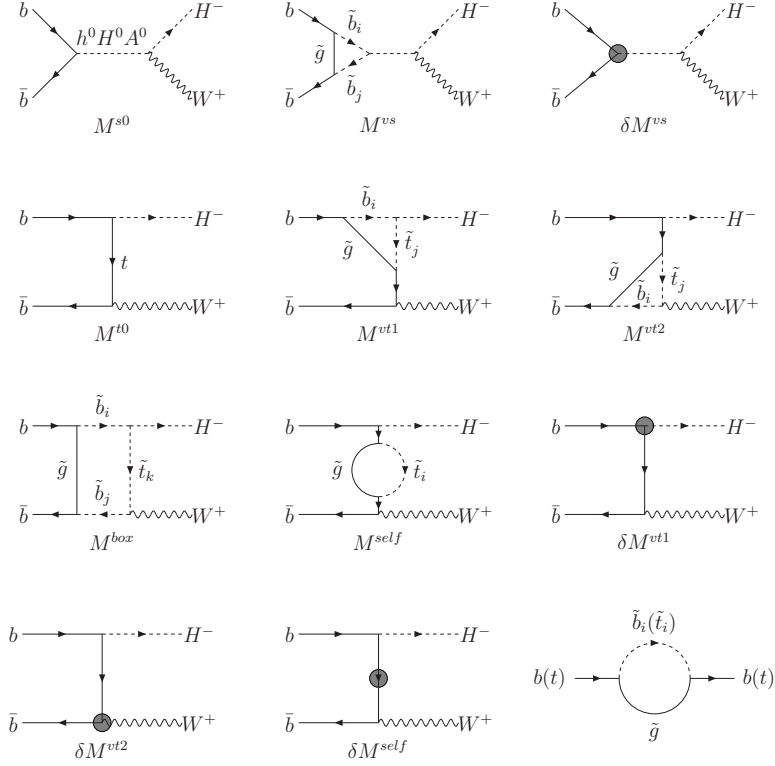,width=400pt}} \vspace{-7cm}
\caption[]{Feynmann diagrams for the subprocess
$b\bar{b}\rightarrow W^{+}H^{-}$. Born diagrams:
$M^{s0}$\,,$M^{t0}$; Virtual correction diagrams:
$M^{vs},M^{vt1},M^{vt2},M^{box},M^{self}$; Counter-term diagrams:
$\delta M^{vs},\delta M^{vt1},\delta M^{vt2},\delta M^{self}$.
\label{feynman}}
\end{figure}

\begin{figure}[h!]
\centerline{\epsfig{file=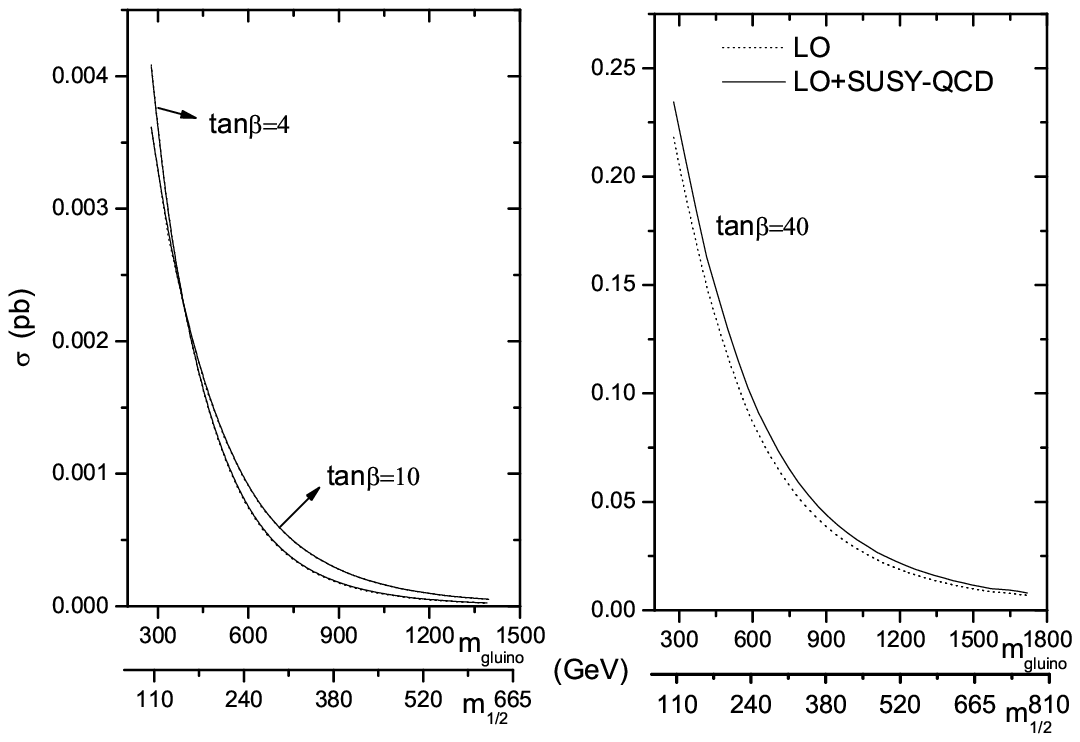,width=400pt}} \caption[]{Total
cross sections for the $W^+H^-$ production at the LHC as functions
of  $m_{\mbox{gluino}}$ or $m_{1/2}$ for $\tan\beta=4\, ,10$ and
$40$, respectively, assuming: $m_0=200$\,\,GeV, $A_0=250$ \,\,GeV,
and $\mu <0$.
 \label{mch}}
\end{figure}

\begin{figure}[h!]
\centerline{\epsfig{file=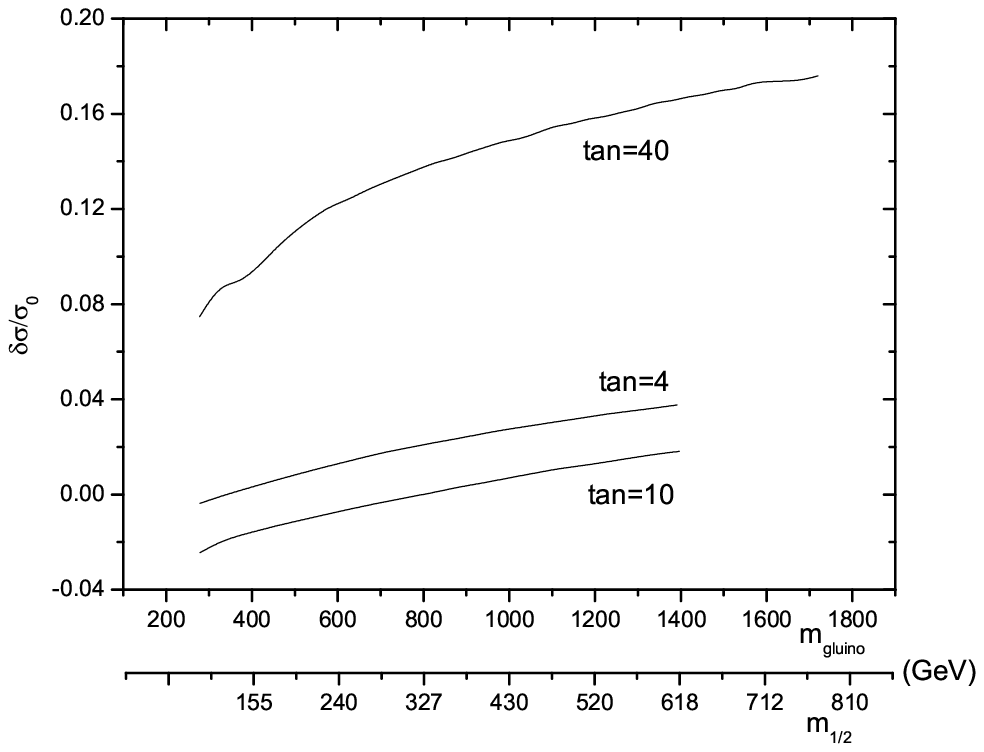,width=400pt}} \caption[]{The
SUSY-QCD relative corrections to the cross sections for the
$W^+H^-$ production at the LHC as functions of $m_{\mbox{gluino}}$
or $m_{1/2}$ for tan$\beta=4\, ,10$ and 40, respectively,
assuming: $m_0=200$\,\,GeV, $A_0=250$ \,\,GeV, and $\mu <0$.
 \label{mch}}
\end{figure}

\begin{figure}[h!]
\centerline{\epsfig{file=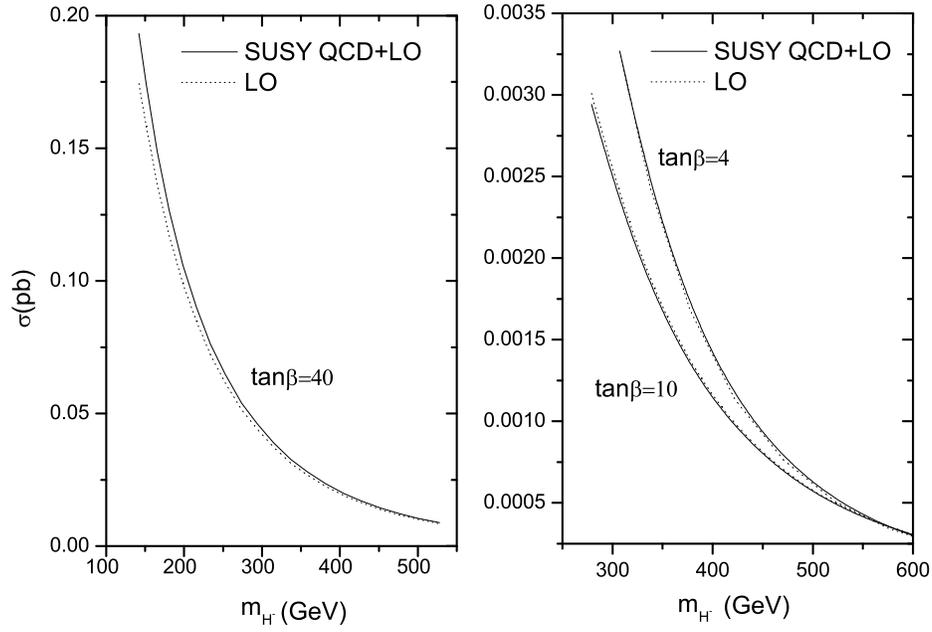,width=400pt}} \caption[]{Total
cross sections for the $W^+H^-$ production at the LHC as functions
of $m_{H^+}$ for tan$\beta=4\, ,10\,$ and 40, respectively,
assuming: $m_{1/2}=180$\,\,GeV, $A_0=250$ \,\,GeV., and $\mu <0$.
 \label{mch}}
\end{figure}

\begin{figure}[h!]
\centerline{\epsfig{file=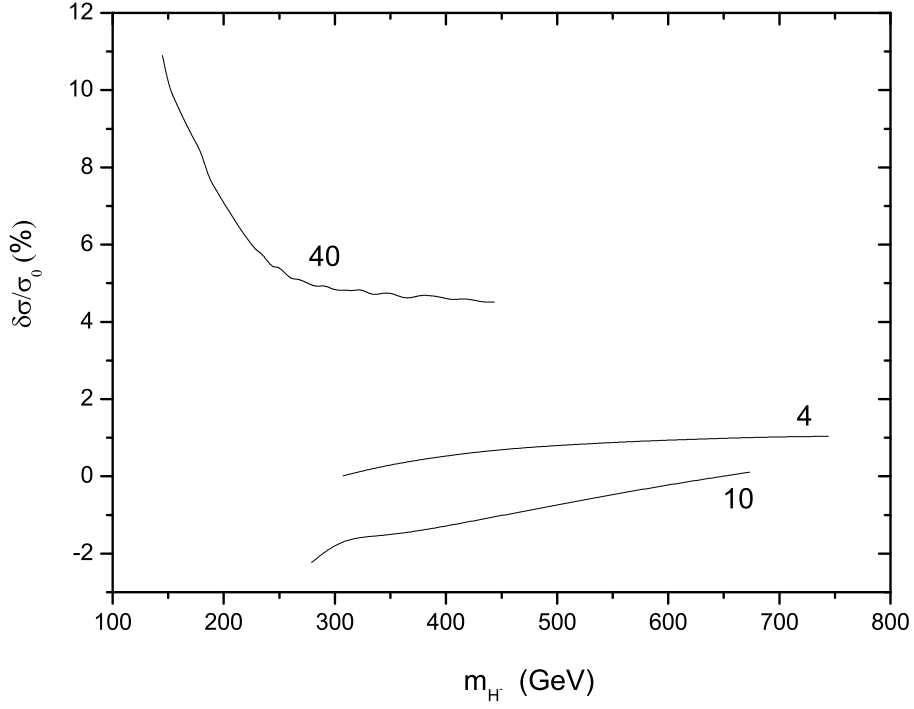,width=400pt}} \caption[]{The
SUSY-QCD relative corrections to the cross sections for the
$W^+H^-$ production at the LHC as functions of $m_{H^+}$ for
tan$\beta=4\, ,10$ and 40, respectively, assuming:
$m_{1/2}=180$\,\,GeV, $A_0=250$ \,\,GeV., and $\mu <0$.
 \label{mch}}
\end{figure}

\begin{figure}[h!]
\centerline{\epsfig{file=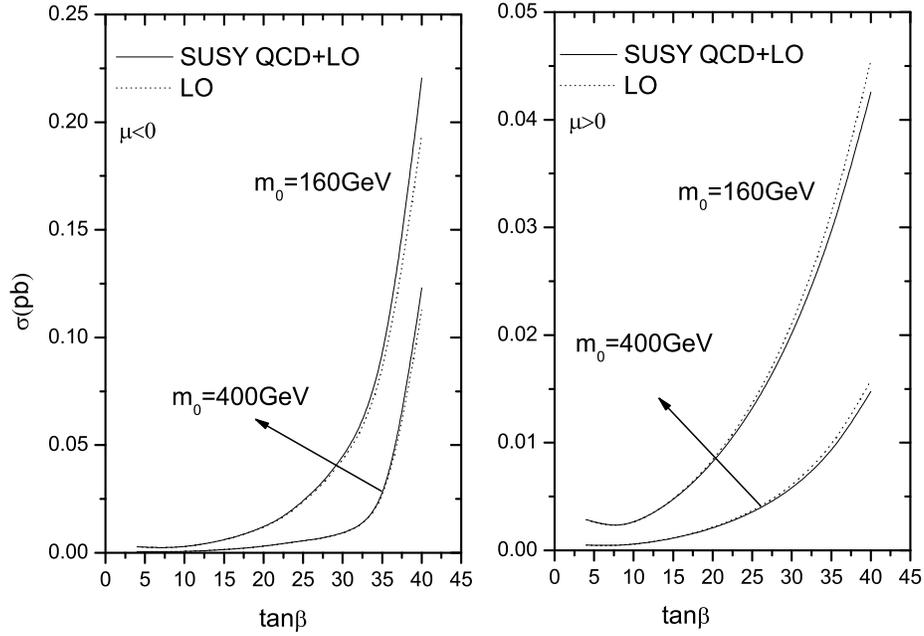,width=400pt}} \caption[]{Total
cross sections for the $W^+H^-$ production at the LHC as a
function of tan$\beta$ for $m_0= 160$ GeV and 400 GeV,
respectively, assuming: $A_0=300$ GeV, $m_{1/2}=160$
GeV.\label{tanb}}
\end{figure}

\begin{figure}[h!]
\centerline{\epsfig{file=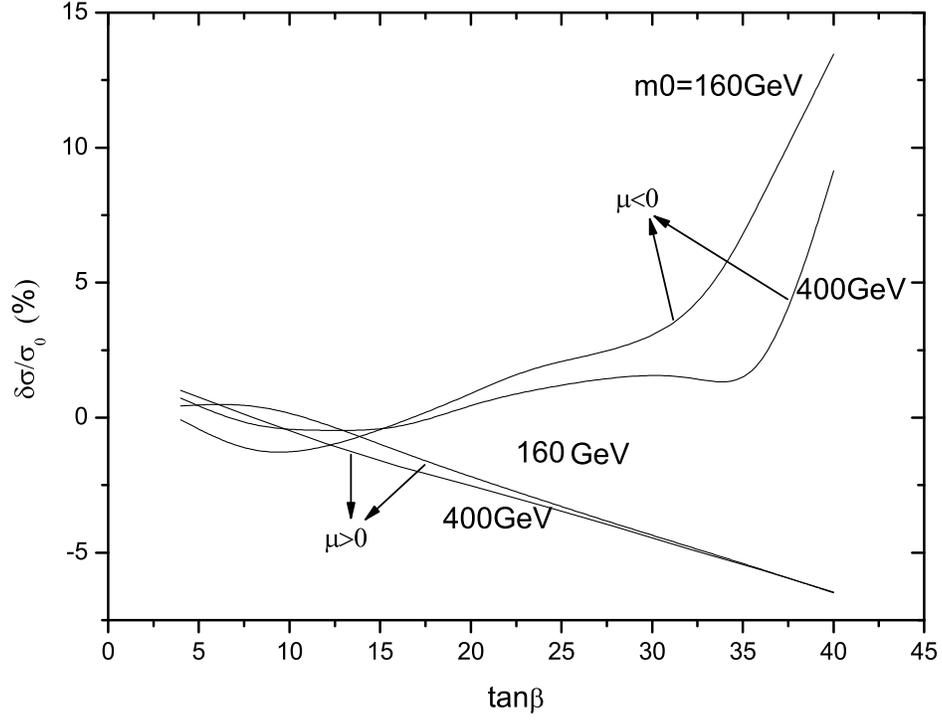,width=400pt}} \caption[]{The
SUSY-QCD relative corrections to the cross sections for the
$W^+H^-$ production at the LHC as a function of tan$\beta$ for
$m_0= 160$ GeV and 400 GeV, respectively, assuming: $A_0=300$ GeV,
$m_{1/2}=160$ GeV.\label{tanb}}
\end{figure}

\begin{figure}[h!]
\centerline{\epsfig{file=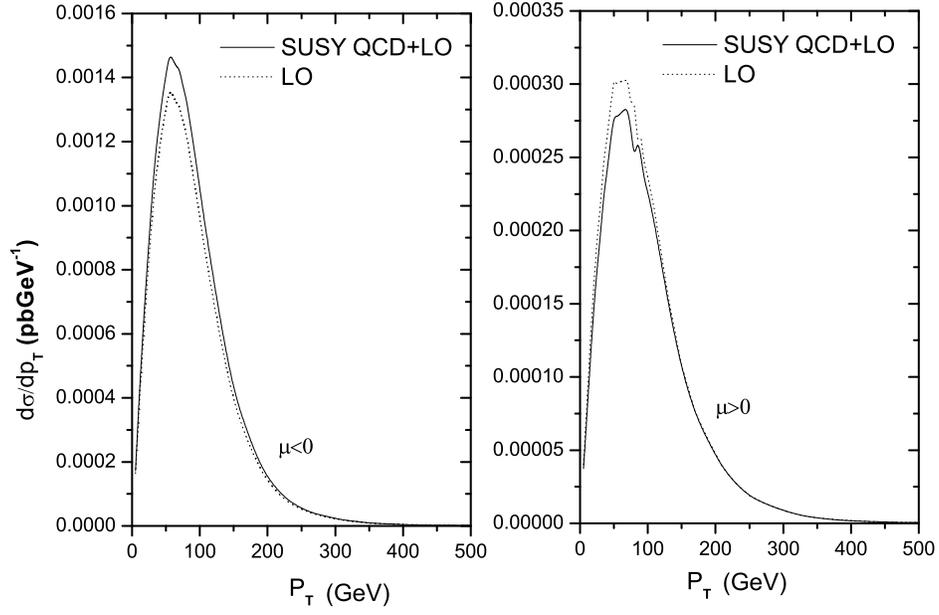,width=400pt}}
\caption[]{Differential cross sections in the transverse momentum
$p_T$ of the W-boson for the $W^+H^-$ production at the LHC,
assuming: $m_0=200$\,\,GeV, $m_{1/2}=180$\,\,GeV,
$A_0=250$\,\,GeV, and $\tan\beta=40.$ \label{y}}
\end{figure}

\begin{figure}[h!]
\centerline{\epsfig{file=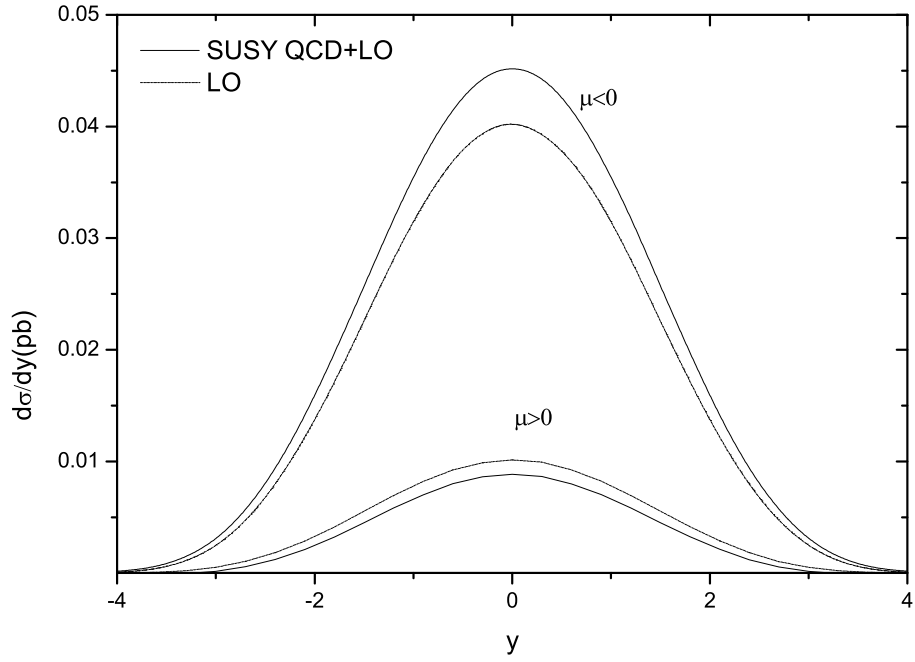,width=400pt}} \caption[]{Differential
cross sections in the rapidity Y of the W-boson for the $W^+H^-$
production at the LHC, assuming: $m_0=200$\,\,GeV,
$m_{1/2}=180$\,\,GeV, $A_0=250$\,\,GeV, and $\tan\beta=40.$
\label{y}}
\end{figure}

\begin{figure}[h!]
\centerline{\epsfig{file=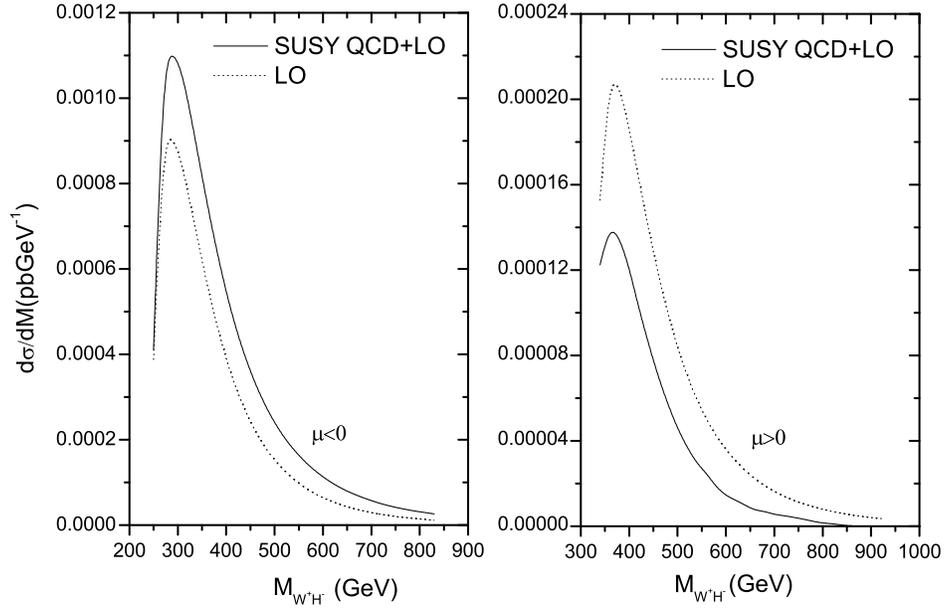,width=400pt}}
\caption[]{Differential cross sections in the invariant mass for
the $W^+H^-$ production at the LHC, assuming: $m_0=200$\,\,GeV,
$m_{1/2}=160$\,\,GeV, $A_0=250$\,\,GeV, and $\tan\beta=40.$
\label{M}}
\end{figure}
\end{document}